\documentclass[aps,prc,twocolumn,10pt,superscriptaddress]{revtex4-1}

\usepackage[utf8]{inputenc}
\usepackage[margin=0.5in]{geometry}
\usepackage{hyperref}
\usepackage{amsmath,amsfonts,amssymb}
\usepackage{graphicx}
\usepackage[dvipsnames]{xcolor}

\hypersetup{
    colorlinks,
    unicode=True,
    linkcolor=Blue,
    citecolor=Blue,
    urlcolor=Blue,
    pdftitle={Transport coefficients of heavy quarks},
    pdfauthor={Ilia Grishmanovskii},
}

\graphicspath{ {figures/} }

\newcommand{\dpi}{(2\pi)}
\newcommand{\qhat}{\hat{q}}
\newcommand{\calO}{\mathcal{O}}
\newcommand{\Msq}[1]{|\overline{\mathcal{M}}_{#1}|^2}
\renewcommand{\vec}{\mathbf}

\begin{document}

\title{Transport coefficients of heavy quarks by elastic and radiative scatterings in the strongly interacting quark-gluon plasma}

\author{Ilia Grishmanovskii}
\email{grishm@itp.uni-frankfurt.de}
\affiliation{Institut für Theoretische Physik, Johann Wolfgang Goethe-Universität,Max-von-Laue-Straße 1, D-60438 Frankfurt am Main, Germany}

\author{Taesoo Song}
\affiliation{GSI Helmholtzzentrum für Schwerionenforschung GmbH, Planckstrasse 1, D-64291 Darmstadt, Germany}

\author{Carsten Greiner}
\affiliation{Institut für Theoretische Physik, Johann Wolfgang Goethe-Universität,Max-von-Laue-Straße 1, D-60438 Frankfurt am Main, Germany}
\affiliation{Helmholtz Research Academy Hesse for FAIR (HFHF), GSI Helmholtz Center for Heavy Ion Physics, Campus Frankfurt, 60438 Frankfurt, Germany}

\author{Elena Bratkovskaya}
\affiliation{GSI Helmholtzzentrum für Schwerionenforschung GmbH, Planckstrasse 1, D-64291 Darmstadt, Germany}
\affiliation{Institut für Theoretische Physik, Johann Wolfgang Goethe-Universität,Max-von-Laue-Straße 1, D-60438 Frankfurt am Main, Germany}
\affiliation{Helmholtz Research Academy Hesse for FAIR (HFHF), GSI Helmholtz Center for Heavy Ion Physics, Campus Frankfurt, 60438 Frankfurt, Germany}

\date{\today}

\begin{abstract}
    We extend our investigation of heavy quark transport coefficients in the effective dynamical quasiparticle model (DQPM) — which reproduces nonperturbative QCD phenomena in the strongly interacting quark-gluon plasma (sQGP) according to lattice QCD data — by including inelastic $2 \to 3$ processes with massive gluon radiation, in addition to elastic $2 \to 2$ parton scattering. Both elastic and inelastic reactions are evaluated at leading order using DQPM-based effective propagators and vertices, accounting for all channels and their interferences. Based on the obtained matrix elements, we calculate various observables connected to a charm quark. First, we calculate the total cross section of a charm quark with the medium partons as functions of temperature and collision energy. Second, we obtain the drag $\mathcal{A}$ coefficient and $\qhat$ coefficient of a charm quark as functions of temperature and momentum and also compare our results with those obtained using the Zakharov model for the momentum-dependent strong coupling for the elastic and radiative vertices with a heavy quark, highlighting the importance of the choice of the strong coupling in the determination of transport coefficients. Third, we calculate the spatial diffusion coefficient of a charm quark and compare our results with those obtained using other approaches. Finally, we explore the mass dependence of the diffusion coefficient by comparing the results for charm quark, bottom quark, and infinitely-heavy quark. We found that inelastic processes can play a significant role in the determination of the transport coefficients at large transverse momenta, but are strongly suppressed at low transverse momenta.
\end{abstract}

\maketitle

\section{Introduction}

Heavy quarks, such as charm and bottom quarks, serve as valuable probes of the quark-gluon plasma (QGP) created in high-energy nuclear collisions. Their large mass ensures that they are primarily produced from early hard scatterings, and their interactions with the medium are not sufficiently strong to fully thermalize them. Consequently, heavy quarks retain information about the QGP as it evolves, making them a powerful tool for studying the properties of the QGP. At high transverse momentum $p_T$, heavy quarks are sensitive to energy loss mechanisms and can shed light on jet quenching phenomena \cite{Zhang:2003wk,Djordjevic:2013pba,Zhang:2018nie,Du:2018yuf,Qin:2015srf,Cao:2020wlm,Xing:2019xae,Uphoff:2014hza}. At low and intermediate $p_T$, heavy quarks are suited for investigating diffusion, thermalization \cite{Moore:2004tg,Cao:2011et}, and color neutralization processes \cite{Plumari:2017ntm,Song:2018tpv,He:2019vgs,Cho:2019lxb,Cao:2019iqs}.

Perturbative QCD (pQCD) offers a natural framework for analyzing heavy-quark interactions in the QGP at sufficiently high temperatures, assuming a small coupling constant $g$. This approach successfully describes heavy-quark dynamics at large $p_T$ \cite{Xing:2019xae,Liu:2021izt} but falls short in capturing the strongly coupled regime that dominates at low and intermediate $p_T$ \cite{Caron-Huot:2008dyw}. As a result, there is a need for other, nonperturbative, approaches to study heavy quark dynamics in the QGP. These methods include lattice QCD \cite{Banerjee:2011ra,Banerjee:2022gen,Brambilla:2020siz,Brambilla:2019tpt,Kaczmarek:2014jga,Francis:2015daa,Ding:2012sp,Altenkort:2023oms,Altenkort:2023eav}, T-matrix approach \cite{Tang:2023tkm,Liu:2016ysz}, effective quasiparticle models \cite{Peshier:2005pp,Cassing:2007nb,Cassing:2007yg,Berrehrah:2016vzw,Moreau:2019vhw,Soloveva:2020hpr,Scardina:2017ipo,Sambataro:2023tlv,Sambataro:2024mkr,Cao:2018ews,Djordjevic:2013xoa}, etc.
 
Useful quantities for comparing various frameworks are transport coefficients, which characterize how heavy quarks interact with and lose energy in the medium. Such transport coefficients include the drag coefficient $\mathcal{A}$, which describes the longitudinal momentum transfer per unit length of the heavy quark, the $\qhat$ coefficient, which describes the transverse momentum transfer per unit length, and the spatial diffusion coefficient $D_s$. The values of these coefficients depend on different factors, including the coupling strength between the heavy quark and the medium, the nature of the plasma, and the underlying dynamics of the interactions \cite{Berrehrah:2016vzw,Cao:2018ews,Rapp:2018qla,Xu:2018gux,Zhao:2023nrz}.

Among the various approaches, the dynamical quasiparticle model (DQPM) \cite{Peshier:2005pp,Cassing:2007nb,Cassing:2007yg,Berrehrah:2016vzw,Moreau:2019vhw,Soloveva:2020hpr} is a promising framework for studying the QGP. The DQPM is constructed to describe the nonperturbative properties of the QGP at finite temperature and baryon chemical potential in terms of strongly interacting quarks and gluons with dynamically generated masses and widths, whose properties are fitted to match the lattice QCD equation of state for the QGP in thermal equilibrium. Since the DQPM incorporates nonperturbative effects, it is well-suited for studying various aspects of heavy-ion collisions, including the properties of the thermal medium \cite{Berrehrah:2014kba,Soloveva:2020hpr,Moreau:2019vhw,Grishmanovskii:2023gog}, jet quenching \cite{Grishmanovskii:2022tpb,Grishmanovskii:2024gag}, and heavy quark dynamics \cite{Berrehrah:2013mua,Berrehrah:2016led,Berrehrah:2016vzw,Song:2019cqz,Song:2022wil}. Although the results were in good agreement with other approaches, the calculations were limited only to elastic $2 \to 2$ processes, and the inclusion of radiative processes was required for a more complete description of heavy quark dynamics in the QGP, especially at large $p_T$.

Recently, we extended the DQPM by explicitly calculating inelastic reactions with  massive gluon emission and applied it to study the properties of the thermal medium \cite{Grishmanovskii:2023gog} and  jet transport coefficients \cite{Grishmanovskii:2024gag}. We showed that, although the inelastic reactions appeared to be insignificant in the context of the thermalized medium due to the predominance of low-energy scatterings, they play an important role when one considers the propagation of a fast jet parton. In particular, we showed a significant impact of the inelastic reactions on the values of $\qhat$ and the energy loss coefficients. We also showed that the jet transport coefficients are highly sensitive to the choice of the strong coupling in different vertices.

In the present work, we aim to explore the impact of radiative processes with the massive gluon emission on the heavy quark dynamics within the QGP. To this aim, we investigate the heavy quark cross section and transport coefficients, particularly drag $\mathcal{A}$ and $\qhat$ coefficients, focusing on the comparison of elastic and inelastic contributions. We also investigate the dependence of the transport coefficients on the choice of the strong coupling used in the vertices connected to the heavy quark and emitted gluon. For this goal, we select the momentum-dependent coupling from the Zakharov model \cite{Zakharov:2020whb,Zakharov:2020psr} and compare the results on the transport coefficients with those obtained using the default temperature-dependent DQPM strong coupling. Furthermore, we calculate the spatial diffusion coefficient $D_s$ and compare our results with those obtained in other approaches. Finally, we explore the mass dependence of the diffusion coefficient by comparing the results for charm quark, bottom quark, and infinitely-heavy quark.

Our study is important for consistent description of charm and bottom dynamics within the microscopic transport approach PHSD (Parton-Hadron-String Dynamics) \cite{Song:2015sfa,Song:2015ykw,Song:2016rzw}, where the QGP phase is modeled based on the DQPM, by explicit accounting for gluon radiative processes additionally to the elastic scattering of heavy quarks with thermal partons.

The paper is organized as follows. In Sec. \ref{sec:DQPM} we briefly recall the basic ideas of the DQPM. In Sec. \ref{sec:methodology} we describe the framework for the calculation of the heavy quark transport coefficients. In Sec. \ref{sec:results} we report on the results for the heavy quark cross sections, drag $\mathcal{A}$ and $\qhat$ coefficients and diffusion coefficient. We summarize our study in Sec. \ref{sec:conclusion}.

\section{\label{sec:DQPM}Dynamical Quasiparticle Model}

\subsection{DQPM ingredients}

The Dynamical Quasiparticle Model (DQPM) \cite{Peshier:2005pp,Cassing:2007nb,Cassing:2007yg,Berrehrah:2016vzw,Moreau:2019vhw,Soloveva:2020hpr} is an effective model that describes the QGP in terms of strongly interacting quarks and gluons, whose properties are fitted to match the lattice QCD calculations of the entropy density in thermal equilibrium and at vanishing chemical potential. The quasiparticles in the DQPM are characterized by "dressed" propagators, i.e., single-particle (two-point) Green's functions, which take the form
\begin{equation}
    G^{R}_j (\omega, \vec{p}) = \frac{1}{\omega^2 - \vec{p}^2 - M_j^2 + 2 i \gamma_j \omega}
    \label{eq:propdqpm}
\end{equation}
for quarks, antiquarks, and gluons ($j = q,\bar q,g$), where $\omega = p_0$ represents energy, $\gamma_{j}$ denote widths, and $M_{j}$ denote thermal masses.

The spectral function of off-shell quasiparticles in the DQPM are parametrized in Lorentzian form with a finite width $\gamma_{j}$ \cite{Linnyk:2015rco}:
\begin{multline}
    \rho_{j}(\omega,\vec{p}) = \frac{\gamma_{j}}{\tilde{E}_j}
    \left(\frac{1}{(\omega-\tilde{E}_j)^2+\gamma^{2}_{j}} - \frac{1}{(\omega+\tilde{E}_j)^2+\gamma^{2}_{j}}\right) 
    \\
    \equiv \frac{4\omega\gamma_j}{\left( \omega^2 - \vec{p}^2 - M^2_j \right)^2 + 4\gamma^2_j \omega^2},
    \label{eq:spectral_function}
\end{multline}
with $\tilde{E}_{j}^2(\vec{p})=\vec{p}^2+M_{j}^{2}-\gamma_{j}^{2}$. The spectral function is antisymmetric in $\omega$ and normalized as
\begin{equation}
    \int\limits_{-\infty}^{\infty}\frac{d\omega}{2\pi}\
    \omega \ \rho_{j}(\omega,\vec{p})=
    \int\limits_{0}^{\infty} d\omega \frac{\omega}{\pi}\ 
    \rho_{j}(\omega,\vec{p})=1.
    \label{eq:spectral_function_norm}
\end{equation}

The DQPM introduces an ansatz (assumption) for the (pole) masses $M_{j}(T,\mu_q)$ and widths $\gamma_{j}(T,\mu_q)$ of quasiparticles as functions of the temperature $T$ and the quark chemical potential $\mu_q$. The pole masses are given by the HTL thermal mass in the asymptotic high-temperature regime -- cf. \cite{Bellac:2011kqa,Linnyk:2015rco} -- for gluons by
\begin{equation}
    M^2_{g}(T,\mu_q)=\frac{g^2(T,\mu_q)}{6}\left(\left(N_{c}+\frac{1}{2}N_{f}\right)T^2
    +\frac{N_c}{2}\sum_{q}\frac{\mu^{2}_{q}}{\pi^2}\right),
    \label{eq:Mg}
\end{equation}
and for quarks (antiquarks) by
\begin{equation}
    M^2_{q(\bar q)}(T,\mu_q)=\frac{N^{2}_{c}-1}{8N_{c}}g^2(T,\mu_q)\left(T^2+\frac{\mu^{2}_{q}}{\pi^2}\right),
    \label{eq:Mq}
\end{equation}
where $N_{c}\; (=3)$ stands for the number of colors, and $N_{f}\; (=3)$ denotes the number of light flavors. We note that Eq. \eqref{eq:Mq} determines the pole masses for the ($u,d$) quarks. The strange quark has a larger bare mass for controlling the strangeness ratio in the QGP. Empirically, we found $M_s(T,\mu_B)= M_{u/d}(T,\mu_B)+ \Delta M$, where $\Delta M \simeq 30$ MeV has been fixed once in comparison to experimental data \cite{Moreau:2019vhw}.

The widths $\gamma_j$ of quasiparticles are taken in the form \cite{Linnyk:2015rco}
\begin{equation}
    \gamma_{j}(T,\mu_\mathrm{B}) = \frac{1}{3} C_j \frac{g^2(T,\mu_\mathrm{B})T}{8\pi}\ln\left(\frac{2c_m}{g^2(T,\mu_\mathrm{B})}+1\right),
    \label{eq:widths}
\end{equation}
where $c_m = 14.4$ is an additional parameter related to a magnetic cutoff, while $C_q = (N_c^2 - 1)/(2 N_c) = 4/3$ and $C_g = N_c = 3$ are the QCD color factors for quarks and gluons, respectively. We also assume that all quarks have the same value of the width. 

Another crucial quantity of the DQPM is the strong coupling, which defines the strength of the interaction between partons and enters the definition of the DQPM masses and widths. In the DQPM, the value of $g^2$ is extracted from lQCD by utilizing a parametrization method introduced in Ref. \cite{Berrehrah:2015vhe}, where it has been shown that for a given value of $g^2$, the ratio $s(T,g^2)/T^3$ is almost constant for different temperatures, i.e., ${\frac{\partial}{\partial T}} (s(T,g^2)/T^3)=0$. Therefore, the entropy density $s$ and the dimensionless equation of state in the DQPM is a function of the effective coupling only, i.e., $s(T,g^2)/s_{SB}(T) = f(g^2)$, where $s^{QCD}_{SB} = 19/9 \pi^2T^3$ is the Stefan-Boltzmann entropy density. Thus, by inverting the $f(g^2)$ function, the coupling $g^2$ can be directly obtained from the parametrization of lQCD data for the entropy density $s(T,\mu_B=0)$ at zero baryon chemical potential:
\begin{equation}
    g^2(T,\mu_B=0) = d \left( \left(s(T,0)/s^\mathrm{QCD}_{SB}\right)^e - 1 \right)^f.
    \label{eq:coupling_DQPM}
\end{equation}
where $s^\mathrm{QCD}_{SB} = 19/9 \pi^2T^3$ is the Stefan-Boltzmann entropy density, and $d = 169.934, e = -0.178434$, and $f = 1.14631$ are the dimensionless parameters obtained by adjusting the quasiparticle entropy density $s(T,\mu_B=0)$ to the lQCD data provided by the BMW Collaboration \cite{Borsanyi:2012cr,Borsanyi:2013bia}. The extension of the strong coupling to finite baryon chemical potential $\mu_B$ is realized using a scaling hypothesis \cite{Cassing:2008nn}, which works up to $\mu_B \approx 500$ MeV. Since the coupling $g^2$ in the DQPM accounts for nonperturbative effects, it appears to be larger compared to the analytical two- or one-loop coupling \cite{Caswell:1974gg}, especially when approaching low temperatures.

We note that, although the baryon chemical potential can affect the partonic cross sections and transport coefficients, its dependence is rather weak compared to the temperature and $\sqrt{s}$ (invariant energy of colliding partons) dependence (see Refs. \cite{Moreau:2019vhw, Grishmanovskii:2023gog}). Moreover, the $\mu_B$ dependence is not considered by other models presented in the paper. Thus, for simplicity and consistency, we only show the results for $\mu_B = 0$ in the present work.

Overall, the DQPM provides the quasiparticle properties, dressed propagators, and the strong coupling, which can be used to evaluate the scattering amplitudes and thus the cross sections and the transport coefficients of quarks and gluons in the QGP as a function of temperature and chemical potential -- cf. Refs. \cite{Berrehrah:2013mua,Moreau:2019vhw,Grishmanovskii:2023gog}. The details of the calculation of the elastic amplitudes and the cross sections are given in Ref. \cite{Moreau:2019vhw}, and for inelastic amplitudes and cross sections in Ref. \cite{Grishmanovskii:2023gog}.

\subsection{Heavy quarks in the DQPM}

Heavy quarks in the DQPM are treated as external probes, which means that they are not part of the thermalized medium. Therefore, we implement a different treatment for the interaction of heavy quarks compared to that of light quarks. We assume that the masses of the heavy quarks are fixed and do not depend on the temperature and chemical potential. Within this study, we use the values $M_c = 1.5$ GeV for charm quarks and $M_b = 4.8$ GeV for bottom quarks. Since the (anti)charm and (anti)bottom quarks are treated as on-shell particles with fixed masses and zero widths, their interactions with other quarks are not self-consistently defined. To address this issue, we introduce an adjustment: for selected heavy quark masses, we employ a modified value of the DQPM strong coupling $g^2(T,\mu_B) \to \sqrt{2} g^2(T,\mu_B)$ at the vertices involving heavy quarks. This modification ensures consistency with lQCD data on $D_s$ as well as with experimental data on charm observables within the PHSD microscopic transport approach \cite{Song:2016rzw}.

Apart from that, we assume that the scattering diagrams for the interaction processes between the bulk and the heavy quarks are the same as for the medium quarks, and their calculation proceeds analogously.

\section{\label{sec:methodology}Methodology}

The propagation of a heavy parton through the thermalized medium can be characterized by transport coefficients, which can be calculated within kinetic transport theory \cite{Svetitsky:1987gq,GolamMustafa:1997id,Moore:2004tg,Berrehrah:2014kba} by accounting for the sequence of elastic and inelastic interactions. 

For elastic ($2\to 2$) reactions the general expression for a transport coefficient reads:
\begin{multline}
    \langle \calO \rangle^{\mathrm{el}} = \frac{1}{2E_j}\sum_{i = q,\bar{q},g}
    \int\frac{d^3p_i}{\dpi^3 2E_i} d_i f_i
    \int\frac{d^3p_1}{\dpi^3 2E_1}
    \\ \times
    \int\frac{d^3p_2}{\dpi^3 2E_2}
    (1 \pm f_1)(1 \pm f_2)
    \\ \times
    \calO \; \Msq{ji \to 12} \; (2\pi)^4 \delta^{(4)}(p_j + p_i - p_1 - p_2),
    \label{eq:TCe_on}
\end{multline}
where $p_i$ is the 4-momentum of the incoming medium parton; $\Msq{ji \to 12}$ is the squared averaged matrix element of the corresponding $(ji \to 12)$ process; $p_1$ and $p_2$ are the outgoing heavy quark and medium parton 4-momenta, respectively; $d_i$ is the medium parton's degeneracy factor for spin and color ($2N_c$ for quarks and $2(N_c^2-1)$ for gluons); $f_i = f_i(E_i,T,\mu_q)$ are the Fermi distribution functions for quarks, and $f_i = f_i(E_i,T)$ are the Bose distribution functions for gluons.

In case of the inelastic ($2 \to 3$) reaction the expression for heavy quark transport coefficients takes the form:
\begin{multline}
    \langle \calO \rangle^{\mathrm{inel}} = \frac{1}{2E_{j}}\sum_{i = q,\bar{q},g}
    \int\frac{d^3p_i}{\dpi^3 2E_i} d_i f_i
    \int\frac{d^3p_1}{\dpi^3 2E_1}
    \\ \times
    \int\frac{d^3p_2}{\dpi^3 2E_2}
    \int\frac{d^3p_3}{\dpi^3 2E_3} 
    (1 \pm f_1)(1 \pm f_2)(1 \pm f_3)
    \\ \times
    \calO \; \Msq{ji \to 123} \; (2\pi)^4 \delta^{(4)}(p_{j} + p_i - p_1 - p_2 - p_3),
    \label{eq:TCi}
\end{multline}
where $\Msq{ji \to 123}$ denotes the squared averaged matrix element of the corresponding radiative $(ji \to 123)$ process, and $p_3$ denotes the momentum of the emitted gluon.

Depending on the choice of $\calO$ in equation \eqref{eq:TCe_on}, one can refer to different transport coefficients:
\begin{itemize}
    \item $\calO = 1$ -- scattering rate $\Gamma$,
    \item $\calO = |p_{j,T} - p_{1,T}|^2$ -- transverse momentum transfer squared $\qhat$ per unit length,
    \item $\calO = E_j - E_1$ -- energy loss $dE/dx$ per unit length,
    \item $\calO = p_{j,L} - p_{1,L}$ -- drag coefficient $\mathcal{A}$.
\end{itemize}
Here, $E_j$, $p_{j,T}$, and $p_{j,L}$ denote the initial (before the collision) values of energy, transverse momentum and longitudinal momentum, respectively, while $E_1$, $p_{1,T}$, and $p_{1,L}$ denote the final (after the collision) values.

We note that Eqs. \eqref{eq:TCe_on} and \eqref{eq:TCi} are formulated for the on-shell case, where the incoming and outgoing medium partons follow the on-shell dispersion relation. In the DQPM, we also consider the off-shell case, where additional integrations over the spectral functions of the medium partons are performed. This integration generally leads to a slight reduction of the scattering cross sections and thus to the transport coefficients. The details of the off-shell calculations for the elastic processes for heavy quarks are given in Ref. \cite{Berrehrah:2014kba}. For inelastic reactions the off-shell calculations are not yet available, and we restrict ourselves to the on-shell case.

\begin{figure*}[t!]
    \centering
    \includegraphics[width=0.9\textwidth]{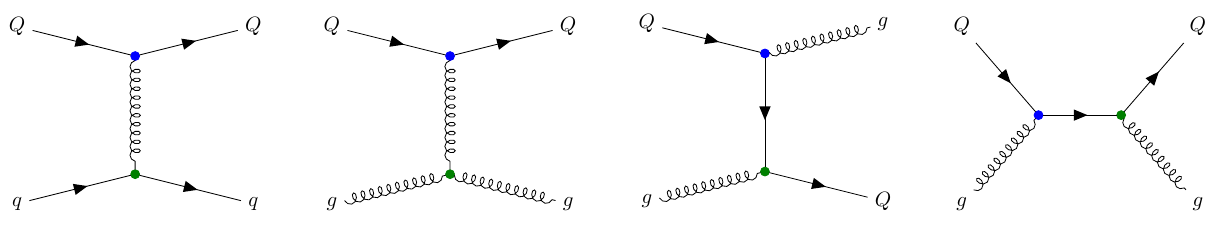}\vspace{0.7cm}
    \includegraphics[width=0.48\textwidth]{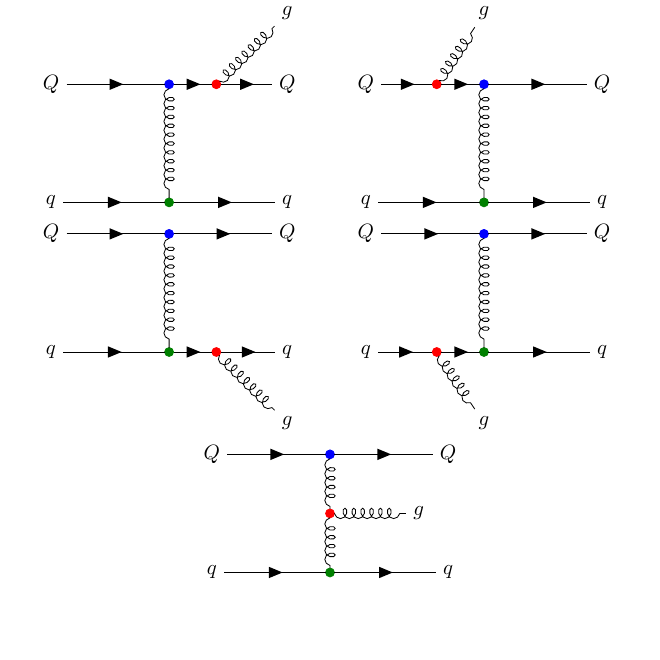}\hfill 
    \includegraphics[width=0.48\textwidth]{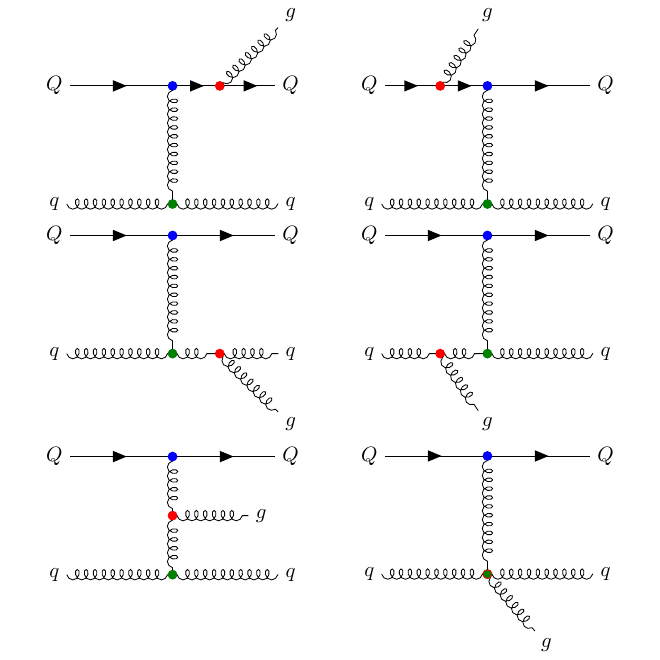}
    \caption{
        Feynman diagrams for the $Q + q \to Q + q$ and $Q + g \to Q + g$ processes (upper) and for the $t$ channel of the $Q + q \to Q + q + g$ (lower left) and $Q + g \to Q + g + g$ (lower right) processes. The symbol $Q$ denotes heavy quark, while $q$ and $g$ denote thermal quark and gluon, respectively. 
        The green dots indicate the vertices corresponding to thermal partons, the blue dots indicate the vertices corresponding to the heavy quark, and the red dots denote the vertices corresponding to the emitted gluon.
    }
    \label{fig:diagrams}
\end{figure*}

The elastic and inelastic scattering diagrams for the interaction between a heavy quark and medium partons are shown in Fig. \ref{fig:diagrams}. For every possible interaction vertex, one can define the corresponding strong coupling: one associated with the thermal parton (green), one associated with the heavy quark (blue dots), and one associated with the emitted gluon (red dots) in the case of inelastic processes. For the ``default" DQPM, the strong coupling in every vertex is defined as the temperature-dependent DQPM coupling given by Eq. \eqref{eq:coupling_DQPM}.

We note that the matrix elements for all processes are calculated explicitly with the use of the DQPM propagators, without any approximations regarding the particle momenta. The corresponding details of the calculation for the elastic amplitudes are given in Ref. \cite{Berrehrah:2016vzw}, and for inelastic amplitudes are given in Ref. \cite{Grishmanovskii:2023gog}.

In the DQPM, the strong coupling is defined as a temperature-dependent coupling, which is fitted to reproduce the lattice QCD data in thermal equilibrium. However, since a heavy quark is not part of the thermal medium, it would be more realistic to consider non-thermal strong couplings for the vertex connected to the heavy quark as well as for the emitted gluon. To see how the choice of a strong coupling affects the values of transport coefficients, we also calculate the transport coefficients with the momentum-dependent coupling from the Zakharov model~\cite{Zakharov:2020whb,Zakharov:2020psr}. There, the strong coupling is defined as
\begin{equation}
    g^2(Q^2) = 
    \begin{cases}
        4\pi \alpha_s^{fr} 
        & \mathrm{if\ } Q \le Q_{fr}, \\
        \frac{48\pi^2}{(11N_c-2N_f)\ln{\left (Q^2/\Lambda^2_{\mathrm{QCD}}\right )}}
        & \mathrm{if\ } Q > Q_{fr},
    \end{cases}
    \label{eq:coupling_Zakharov}
\end{equation}
with $\Lambda_{\mathrm{QCD}} = 0.2$ GeV, $Q_{fr} = \Lambda_{\mathrm{QCD}} \exp(2\pi/9\alpha_s^{fr})$, and $\alpha_s^{fr}$ is a free parameter. In this work, we consider $\alpha_s^{fr} = 1.05$ (vacuum value). For both elastic and radiative collisions for the heavy quark vertex the value of $Q$ in Eq. \eqref{eq:coupling_Zakharov} is defined as the momentum transfer between the heavy quark and the medium parton. For the vertex connected to the emitted gluon in $2 \to 3$ reaction, the value of $Q$ is suggested to be $k_T$, the transverse momentum of the emitted gluon. For the thermal vertex, the coupling remains to be $g^{\mathrm{DQPM}}(T)$. We note that in this scenario the strong coupling is modified only in the vertices, while the masses and widths of the quasiparticles remain unchanged.

\section{\label{sec:results}Results}

\subsection{\label{sec:cross_section}Heavy quark cross section}

We start by investigating heavy quark interactions with the quark-gluon plasma in terms of scattering cross sections.

\begin{figure*}[ht!]
    \centering
    \includegraphics[width=0.95\columnwidth]{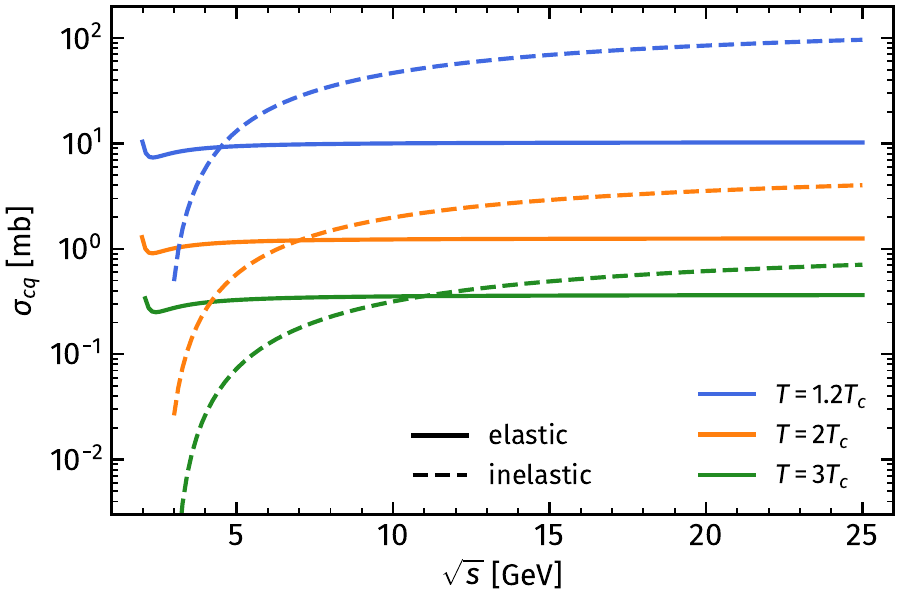}\hspace{1cm}
    \includegraphics[width=0.95\columnwidth]{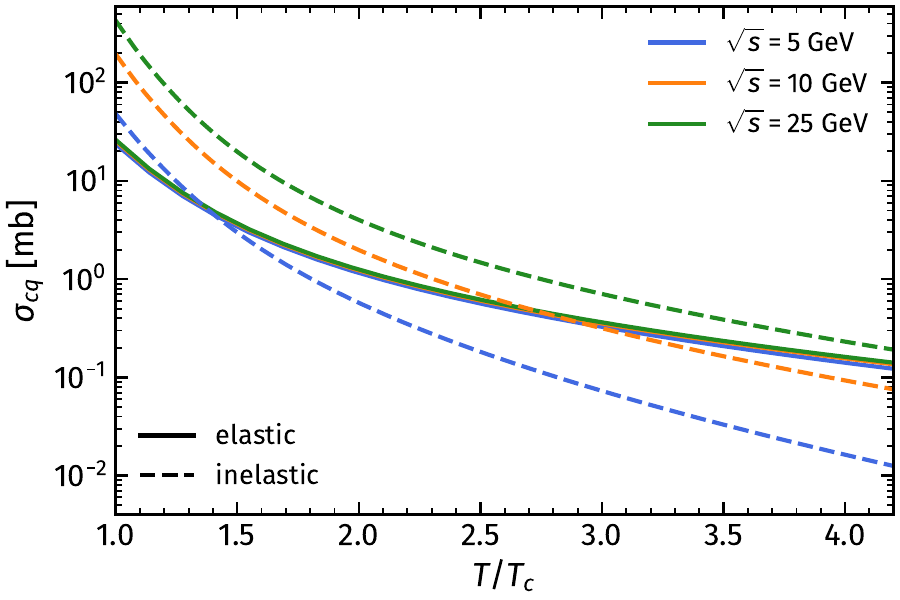}
    \caption{
        Total on-shell elastic (solid lines) and inelastic (dashed lines) cross sections for a charm quark as a function of invariant collision energy $\sqrt{s}$ of $c + q$ scattering at different temperatures (left plot) and as a function of temperature for different collision energies (right plot).
    }
    \label{fig:cross_section}
\end{figure*}

Figure \ref{fig:cross_section} shows the total on-shell elastic ($c + q \to c + q$) and inelastic ($c + q \to c + q + g$) cross sections for a charm quark as a function of invariant collision energy $\sqrt{s}$ of $c + q$ scattering at different temperatures, and as a function of temperature for different collision energies. We do not show the results for the bottom quark, since they are qualitatively similar to those of the charm quark, but slightly differ quantitatively due to the larger mass. We also do not show the results for the interaction of a heavy quark with gluons, implying the scaling $\sigma_{cg} = \frac{9}{4} \sigma_{cq}$, which is valid for both the elastic and inelastic processes (cf. Ref. \cite{Grishmanovskii:2023gog}). We note that in Fig. \ref{fig:cross_section} the cross sections are shown for the default DQPM strong coupling.

While the elastic cross sections grow slowly with increasing energy and approach asymptotic behavior at large $\sqrt{s}$, the inelastic reactions exhibit a strong energy dependence, increasing monotonically over the entire range of scattering energies. The nature of this energy dependence follows from the structure of the scattering amplitudes and does not depend on the strong coupling, since the strong coupling is not momentum-dependent. In contrast, the temperature dependence is governed primarily by the behavior of the DQPM strong coupling. For elastic reactions, the squared amplitudes are proportional to the coupling squared ($\Msq{2\to2} \propto \alpha_s^2$), while for inelastic scatterings the amplitudes are proportional to the coupling cubed ($\Msq{2\to3} \propto \alpha_s^3$). Because of this, the strong temperature dependence of the DQPM strong coupling drives the strong temperature dependence of the total cross sections, with inelastic reactions showing a greater dependence. 

Overall, one can conclude that elastic reactions dominate at low energies and high temperature, while inelastic reactions dominate at high energies and low temperature. This picture suggests that the transport coefficients associated with the heavy quark should show the dominance of elastic reactions at small and intermediate $p_T$ of a heavy quark and the dominance of inelastic reactions at large $p_T$. This is a subject of the next section.

\subsection{Drag and \texorpdfstring{$\qhat$}{qhat} coefficients}

In this section, we investigate the drag $\mathcal{A}$ and $\qhat$ coefficients of a charm quark, which characterize the longitudinal and transverse momentum transfer per unit length, respectively.

\begin{figure*}
    \centering
    \includegraphics[width=0.85\columnwidth]{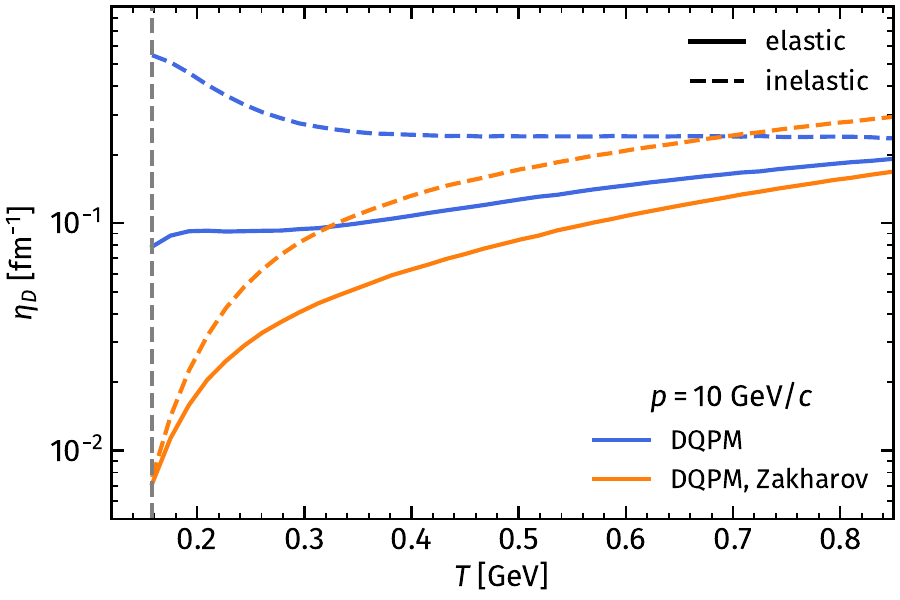}\hspace{0.5cm}
    \includegraphics[width=0.85\columnwidth]{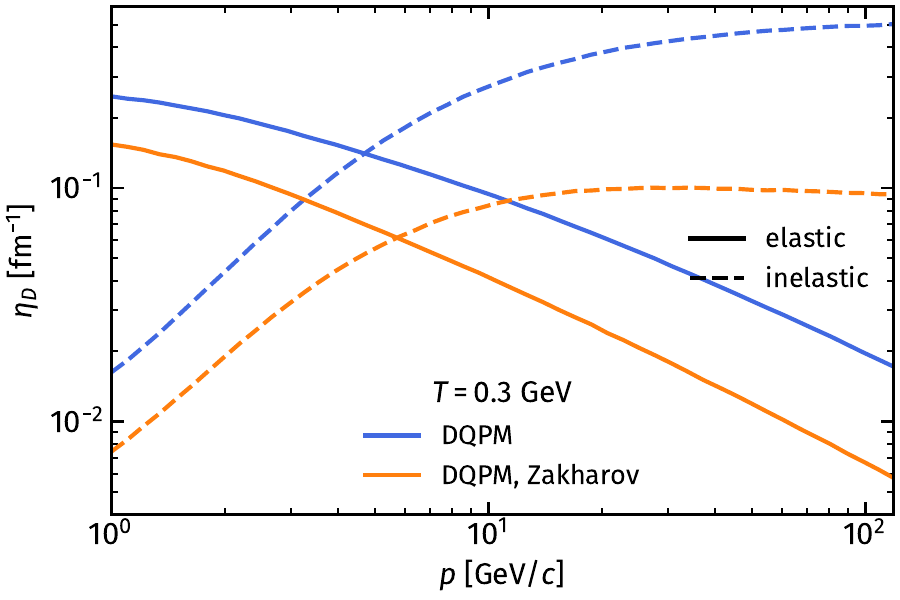}\vspace{0.2cm}
    \includegraphics[width=0.85\columnwidth]{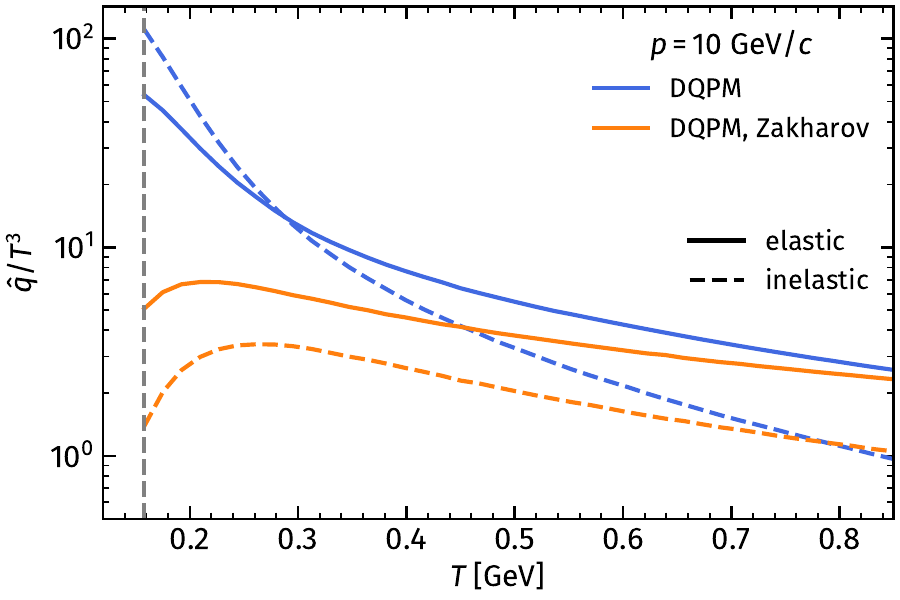}\hspace{0.5cm}
    \includegraphics[width=0.85\columnwidth]{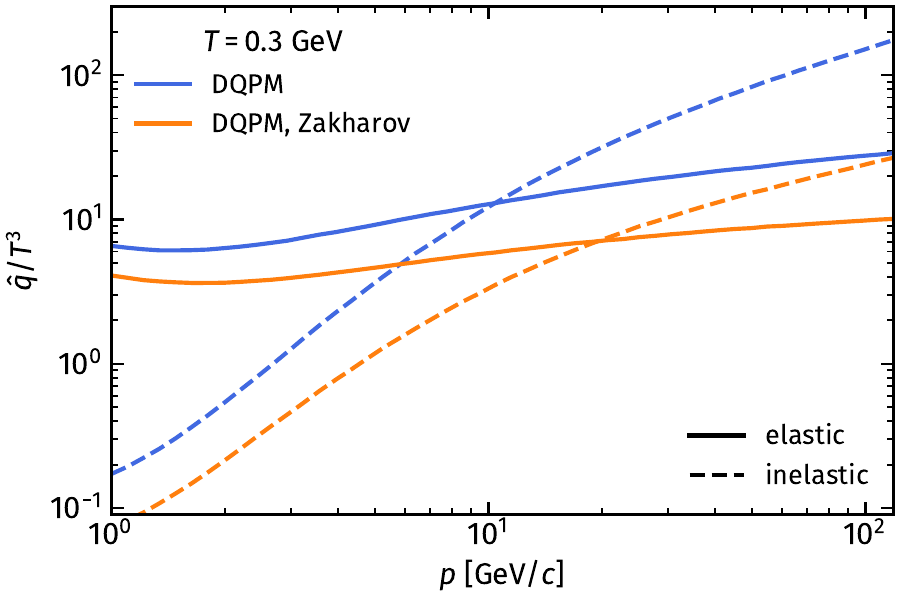}
    \caption{
        Drag (upper plots) and $\qhat$ (lower plots) coefficients of a charm quark as functions of temperature (left plots) and functions of charm momentum (right plots), calculated separately for elastic and inelastic reactions. Blue lines represent the default DQPM results, while orange lines represent the results from the momentum-dependent coupling from the Zakharov model.
    }
    \label{fig:drag_qhat_charm}
\end{figure*}

Figure \ref{fig:drag_qhat_charm} shows the scaled drag $\eta_D = \mathcal{A}/p$ and the $\qhat$ coefficient of a charm quark as functions of temperature at fixed charm quark momentum, and as functions of charm momentum for a fixed temperature, calculated separately for elastic and inelastic reactions. The blue lines on the figure show the default DQPM results, where all strong couplings are taken from the DQPM, while the orange lines represent results with the strong couplings taken from the Zakharov model (see Eq. \eqref{eq:coupling_Zakharov}).

For the default DQPM results, the elastic contribution dominates at higher temperatures and lower charm momenta, whereas the inelastic contribution tends to prevail at lower temperatures and higher charm momenta. This behavior is consistent with the cross-section results discussed previously: at large charm momenta, the average energy of scatterings between the charm quark and medium partons increases, causing inelastic reactions to dominate, while at low momenta the inelastic contribution is strongly suppressed. A similar pattern appears in the temperature dependence: at low temperatures, inelastic reactions become more significant due to the high value of the DQPM strong coupling, and as the temperature rises, their contribution decreases.

When using the DQPM with Zakharov couplings, the results differ both quantitatively and qualitatively, reflecting the different dependence of the strong coupling. While the DQPM strong coupling depends solely on temperature and not on momentum, the Zakharov coupling is momentum-dependent and decreases as momentum increases. Consequently, there is a weaker temperature dependence and a suppression of the transport coefficients at large charm momenta. This also explains the small difference between the elastic and inelastic cases for DQPM with the Zakharov coupling at low temperatures: since the Zakharov coupling does not depend on $T$, it does not contribute to the temperature dependence of the cross sections and thus of the transport coefficients. Although the DQPM and Zakharov-based results tend to converge at high temperatures, they differ significantly at temperatures up to about 0.5 GeV, which corresponds to the typical temperature range of the QGP formed in heavy-ion collisions.

We note that the observed results for a charm quark are qualitatively similar to those obtained for the jet parton (cf. Ref. \cite{Grishmanovskii:2024gag}), but differ quantitatively due to the higher mass of a charm quark and different strong couplings.

\subsection{Diffusion coefficient}

The spatial diffusion coefficient $D_s$ is another significant transport parameter that characterizes the interaction of heavy quarks with the medium, which is directly related to the thermalization time and can be evaluated also in lattice QCD. The spatial diffusion coefficient $D_s$ of a heavy quark is defined as
\begin{equation}
    D_s = \lim_{p \to 0} \frac{T}{(\mathcal{A}/p) M},
    \label{eq:diffusion_coef}
\end{equation}
where $M$ is the mass of the heavy quark.

\begin{figure}[ht!]
    \centering
    \includegraphics[width=\columnwidth]{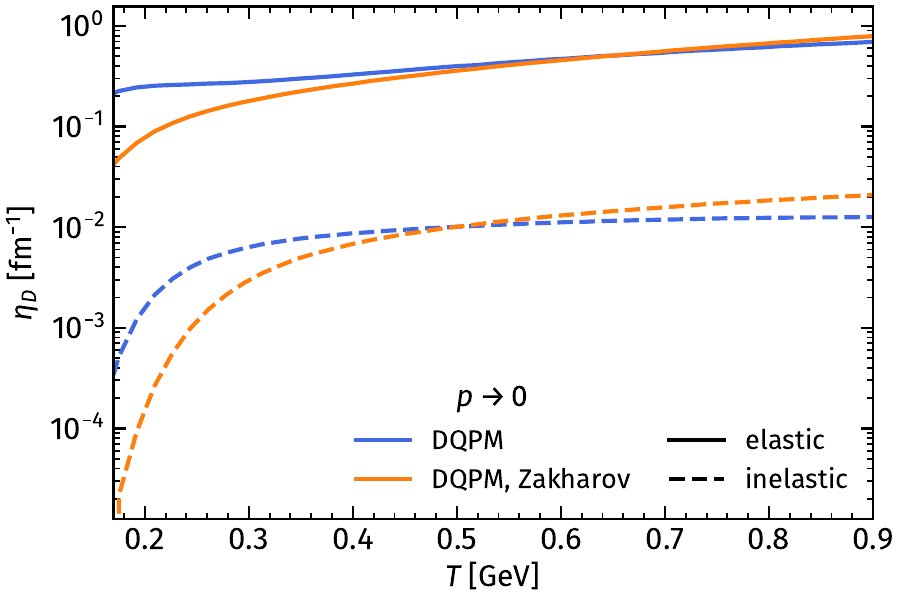}
    \caption{
        Drag coefficient of a charm quark as a function of temperature for $p \to 0$, calculated separately for elastic and inelastic reactions. Blue lines represent the default DQPM results, while orange lines represent the results from the momentum-dependent coupling from the Zakharov model.
    }
    \label{fig:DRAG_T_charm_p0}
\end{figure}

Figure \ref{fig:DRAG_T_charm_p0} shows the scaled drag $\eta_D = \mathcal{A}/p$ coefficient of a charm quark as a function of temperature for $p \to 0$, calculated separately for elastic and inelastic reactions. One can see that the drag coefficient is strongly dominated by elastic scatterings in the entire temperature range both for the DQPM and the Zakharov coupling. However, the Zakharov coupling leads to a significantly smaller value of the drag coefficient at low temperatures, while at high temperatures the results converge.

\begin{figure}[ht!]
    \centering
    \includegraphics[width=\columnwidth]{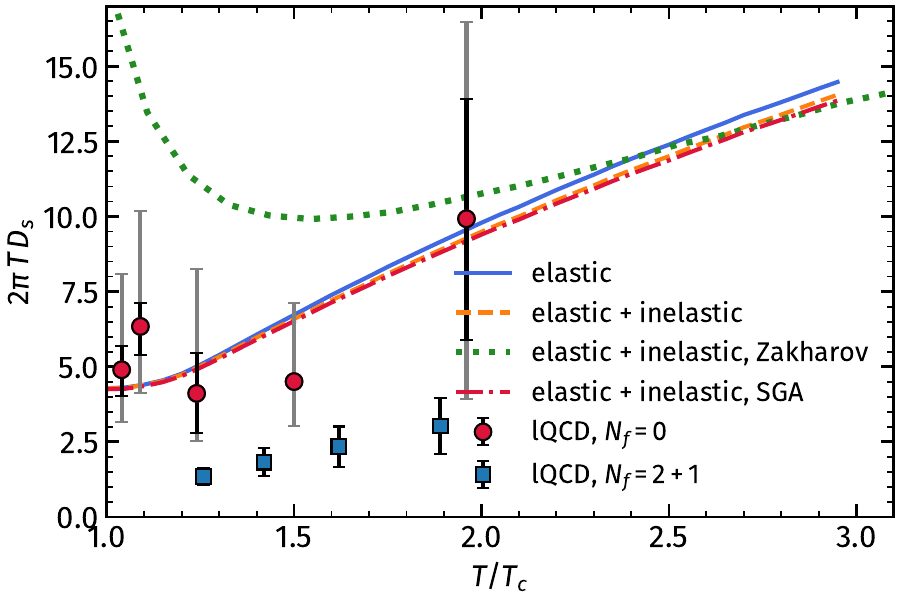}
    \caption{
        Spatial diffusion coefficient $D_s$ of the charm quark. Blue solid line represents results for only elastic scattering. Orange dashed line represents the sum of elastic and radiative scatterings with the thermal emitted gluon. Green dotted line represents the results with Zakharov coupling. Red dash-dotted line represents results for elastic and inelastic scatterings with inelastic amplitudes calculated within the SGA approach from Ref. \cite{Song:2022wil}. Red error bars represent the lQCD data from Ref. \cite{Banerjee:2011ra}, and the blue error bars represent the lQCD data from Ref. \cite{Altenkort:2023eav}.
    }
    \label{fig:Ds_off}
\end{figure}

Figure \ref{fig:Ds_off} illustrates the spatial diffusion coefficient $D_s$ of the charm quark for only elastic scattering and the sum of elastic and radiative scatterings for two different cases: pure DQPM results, results with the Zakharov coupling, and the results with inelastic scatterings calculated within the soft-gluon approximation (SGA) approach from Ref. \cite{Song:2022wil}. We note that the $D_s$ coefficient is calculated in the off-shell mode only for the elastic case, while the inelastic case is calculated in the on-shell mode. To provide an appropriate scale the results from lattice simulations for a quenched QCD \cite{Banerjee:2011ra} and for (2+1)-flavor QCD \cite{Altenkort:2023eav} are also shown. One can see that the inclusion of inelastic reactions with massive emitted gluons only slightly reduces the value of the diffusion coefficient. This behavior is attributed to the fact that the diffusion coefficient is calculated in the low-momentum limit of a heavy quark, where the interaction is dominated by elastic scatterings (see Fig.~\ref{fig:drag_qhat_charm}). It is also seen that the results from the SGA approach are consistent with the full DQPM calculations, showing only minor deviations in the values of the diffusion coefficient. In contrast, the results with the Zakharov coupling show a significant enhancement of the diffusion coefficient at low temperatures, which is attributed to the corresponding reduction of the drag coefficient in Fig. \ref{fig:DRAG_T_charm_p0}. This result should not be interpreted as a physical one, since the Zakharov coupling is constructed primarily to describe a fast jet parton, and the application of this coupling to a slow moving charm quark is not justified.

\begin{figure}[ht!]
    \centering
    \includegraphics[width=\columnwidth]{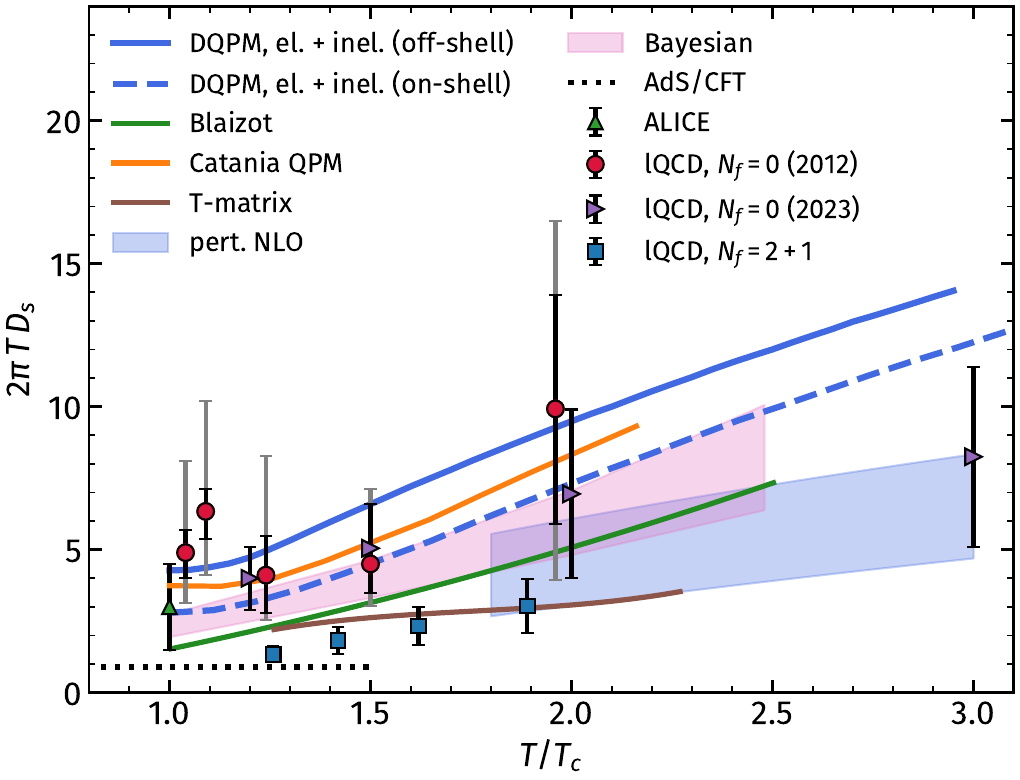}
    \caption{
        DQPM results for the spatial diffusion coefficient for the off-shell and the on-shell cases compared with the results from, quenched lQCD \cite{Banerjee:2011ra,Banerjee:2022gen}, (2+1)-flavor lQCD \cite{Altenkort:2023eav}, the Catania QPM \cite{Sambataro:2023tlv}, T-matrix approach \cite{Tang:2023tkm,Liu:2016ysz}, Blaizot formalism \cite{Blaizot:2015hya}, AdS/CFT estimate \cite{Casalderrey-Solana:2006fio}, ALICE phenomenological estimate \cite{Xu:2017obm,ALICE:2021rxa}, NLO perturbative calculations \cite{Caron-Huot:2007rwy}, and Bayesian analysis by the Duke QCD group \cite{Xu:2017hgt} (see text for details).
    }
    \label{fig:Ds_charm_comparison}
\end{figure}

To provide a more complete overview of the DQPM results, we show in Fig.~\ref{fig:Ds_charm_comparison} a comparison of the DQPM calculations for the spatial diffusion coefficient of a charm quark with various approaches. These include:
\begin{itemize}
    \item the Catania quasiparticle approach for $N_f = 2+1$ \cite{Sambataro:2023tlv,Scardina:2017ipo}, which, similar to the DQPM, is based on the quasiparticle picture of the QGP, but uses a different parametrization for the strong coupling and the mass of the medium partons
    \item the T-matrix approach \cite{Tang:2023tkm,Liu:2016ysz}
    \item the AdS/CFT estimate \cite{Casalderrey-Solana:2006fio}
    \item the ALICE collaboration's phenomenological estimate for an infinitely-heavy quark \cite{Xu:2017obm,ALICE:2021rxa}
    \item the NLO perturbative calculations, obtained at the renormalization scale from $\mu = 2\pi T$ to $\mu = 4\pi T$ \cite{Caron-Huot:2007rwy} in the limit of an infinitely heavy quark
    \item the Bayesian analysis by the Duke QCD group \cite{Xu:2017hgt}, based on calibrating to the experimental data of $D$-meson $R_{AA}$ and $v_2$ in Au+Au collisions at invariant $NN$ energy $\sqrt{s_{NN}} = 200$ GeV and Pb+Pb collisions at $\sqrt{s_{NN}} = 2.76$ TeV
    \item the lattice data for quenched QCD \cite{Banerjee:2011ra,Banerjee:2022gen} and for (2+1)-flavor QCD \cite{Altenkort:2023eav}
    \item the formalism by Blaizot et al. \cite{Blaizot:2015hya}, based on a generalized Langevin equation.
\end{itemize}

One can see that both on-shell and off-shell DQPM results are in good agreement with the quenched lattice data, but overestimate the data from ($2+1$)-flavor lQCD. The DQPM results for the off-shell case also show the highest values of the diffusion coefficient across the entire range of temperatures when compared to the other approaches, even after including inelastic contributions. Nevertheless, the DQPM results are consistent with the other approaches, showing the same trend of increasing diffusion coefficient with increasing temperature.

Now, we turn to the comparison of the mass dependence of the spatial diffusion coefficient. In the DQPM, the difference between a charm quark and a bottom quark is only in the mass, which is fixed and does not depend on temperature. Therefore, the DQPM allows us to study the mass dependence of the diffusion coefficient without any additional assumptions.

\begin{figure}[t!]
    \centering
    \includegraphics[width=\columnwidth]{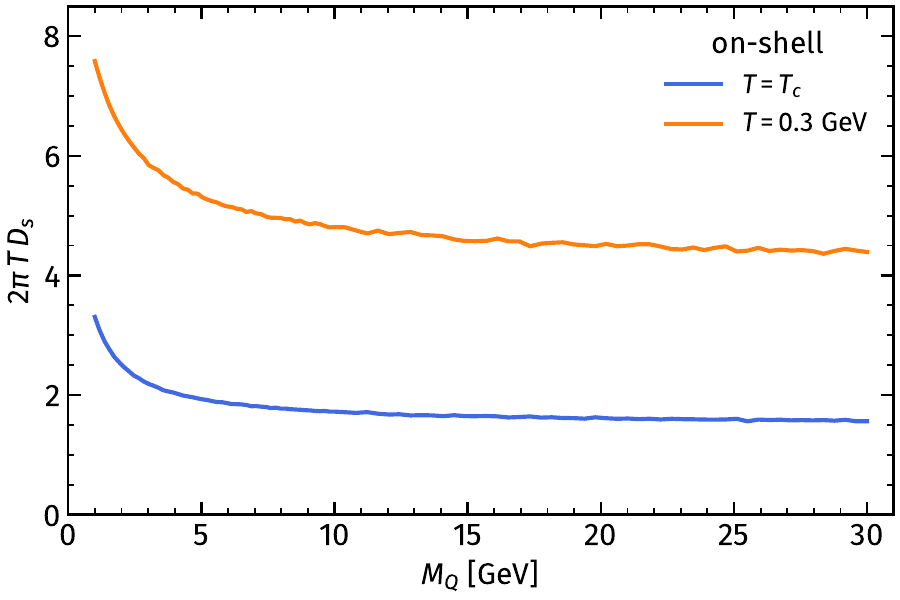}
    \caption{
        DQPM on-shell results for the spatial diffusion coefficient of a heavy quark as a function of the heavy quark mass at $T = T_c$ and $T = 0.3$ GeV.
    }
    \label{fig:Ds_M}
\end{figure}

Figure \ref{fig:Ds_M} illustrates the spatial diffusion coefficient of a heavy quark as a function of the heavy quark mass. The results show a clear mass dependence with the diffusion coefficient decreasing with increasing mass of the heavy quark and saturating at higher masses, regardless of the temperature. Thus, one can achieve the infinitely-heavy quark limit by taking the mass of the heavy quark to be larger than, e.g., 15 GeV. Although the results are presented for the on-shell case, the off-shell case shows a similar mass dependence, but with slightly larger values of the diffusion coefficient.

\begin{figure}[t!]
    \centering
    \includegraphics[width=\columnwidth]{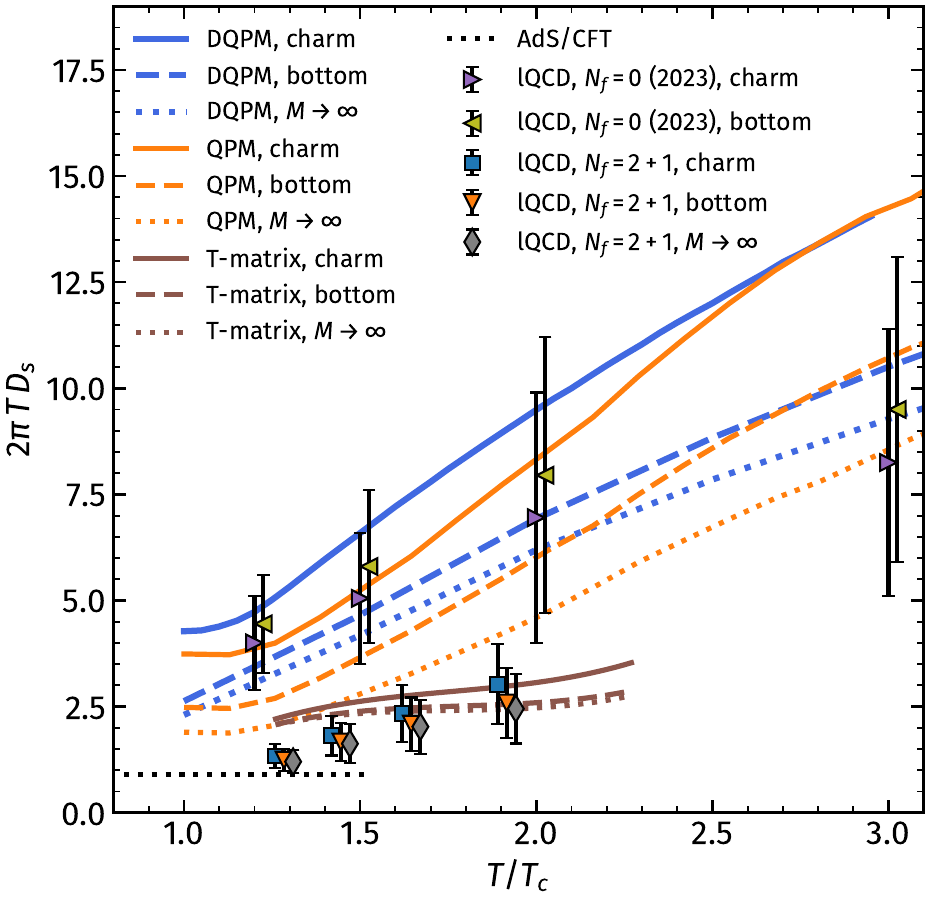}
    \caption{
        DQPM off-shell results for the spatial diffusion coefficient for a charm quark (solid line), bottom quark (dashed line), and for an infinitely heavy quark (dotted line), in comparison with other approaches (see legend in Fig.~\ref{fig:Ds_charm_comparison}).
    }
    \label{fig:Ds_comparison_mass}
\end{figure}

Figure \ref{fig:Ds_comparison_mass} shows the DQPM spatial diffusion coefficient for the off-shell case for a charm quark, bottom quark, and an infinitely heavy quark, in comparison with the results from other approaches, where the data for a bottom quark and an infinitely heavy quark are available. For all presented approaches, except the one in Ref. \cite{Banerjee:2022gen}, the diffusion coefficient slightly decreases with increasing mass of the heavy quark, although the strength of the mass dependence varies from one approach to another.

\section{\label{sec:conclusion}Conclusions}

In this work, we have extended the investigation of heavy-quark transport coefficients within the DQPM by accounting for inelastic $2 \to 3$ reactions, with massive gluon radiation, in addition to the elastic $2 \to 2$ scatterings of partons. The inelastic reactions are calculated explicitly within leading-order Feynman diagrams with effective propagators and vertices from the DQPM by accounting for all channels and their interferences. We calculated the charm quark total cross sections for elastic and inelastic reactions, the corresponding drag and $\qhat$ coefficients, and the spatial diffusion coefficient. The results are compared with the results from other approaches, including the Zakharov model with momentum-dependent coupling. 
Our results can be summarized as follows:
\begin{itemize}
    \item The total cross sections for elastic and inelastic reactions between a heavy quark and the medium show strong energy and temperature dependencies. While the inelastic reactions dominate at high energies and low temperatures, the elastic ones dominate at low energies and high temperatures.
    \item The inelastic reactions play an important role in the determination of the heavy quark transport coefficients, especially at low temperatures and high momenta.
    \item The drag and $\qhat$ coefficients show a high sensitivity to the choice of the strong coupling. The pure DQPM results are generally larger than those obtained with the Zakharov couplings, especially at low temperatures, where the DQPM strong coupling is large.
    \item The inelastic reactions play a minor role in the determination of the spatial diffusion coefficient, which is obtained at low momentum.
    \item The spatial diffusion coefficient $D_s$ of a charm quark from the DQPM is consistent with the other approaches, although it shows larger values across the entire range of temperatures.
    \item The mass dependence of the diffusion coefficient shows a decrease with increasing mass of the heavy quark for the DQPM and for all approaches presented in this work.
\end{itemize}

This study is relevant for the interpretation of heavy quark observables in heavy-ion collisions, such as the nuclear modification factor $R_{AA}$ and elliptic flow $v_2$, which are sensitive to the transport coefficients. This issue can be addressed in future studies by implementing the inelastic scattering processes in the PHSD transport approach.

\section*{ACKNOWLEDGEMENTS}

The authors acknowledge inspiring discussions with J. Aichelin and W. Cassing. Furthermore, we acknowledge support by the Deutsche Forschungsgemeinschaft (DFG, German Research Foundation) through the grant CRC-TR 211 "Strong-interaction matter under extreme conditions" -- project number 315477589 -- TRR 211. This work is supported by the European Union's Horizon 2020 research and innovation program under grant agreement No 824093 (STRONG-2020). The computational resources have been provided by the Goethe-HLR Center for Scientific Computing.


\bibliography{refs}

\end{document}